\documentclass[a4,11pt]{article}
\usepackage{epsfig}
\usepackage{epstopdf}

\usepackage{amssymb,epsfig,graphicx,epsf}

\usepackage[titletoc,toc,title]{appendix}

\usepackage{amsfonts}
\usepackage{amsmath}
\usepackage{feynmf}

\usepackage{latexsym}
\usepackage{float} 

\usepackage{slashed}
\usepackage{cite}
\usepackage[utf8]{inputenc} 
\usepackage{mathtools}
\usepackage[font=small,skip=-28pt]{caption}
\usepackage{color}
\DeclareFontFamily{OT1}{pzc}{}
\DeclareFontShape{OT1}{pzc}{m}{it}{<-> s * [1.10] pzcmi7t}{}
\DeclareMathAlphabet{\mathpzc}{OT1}{pzc}{m}{it}

\usepackage{commath} 
\usepackage{accents} 
\usepackage{relsize} 
\usepackage[makeroom]{cancel} 
\usepackage{xcolor}
\usepackage[normalem]{ulem} 

\newcommand{\be}{\begin{equation}}
\newcommand{\ee}{\end{equation}}

\newcommand{\ba}{\begin{eqnarray}}
\newcommand{\ea}{\end{eqnarray}}
\newcommand{\bal}{\begin{aligned}}
\newcommand{\eal}{\end{aligned}}

\begin{document}

\title{Bose-Einstein graviton condensate in a Schwarzschild black hole}

\author{Jorge Alfaro\\
Pontificia Universidad Cat\'olica de Chile, Av. Vicu\~na Mackenna 4860, Santiago, Chile,\\ 
\\
Dom\`enec Espriu and  Luciano Gabbanelli\\
Departament of Quantum Physics and Astrophysics and\\ 
Institut de Ci\`encies del Cosmos (ICCUB), Universitat de Barcelona\\
Mart\'\i ~i Franqu\`es, 1, 08028 Barcelona, Spain.}

\date{}

\maketitle

\begin{abstract}
We analyze in detail a previous proposal by Dvali and G\'omez that black holes could be treated as
consisting of a Bose-Einstein condensate of gravitons. In order to do so we extend the Einstein-Hilbert action with a 
chemical potential-like term, thus placing ourselves in a grand-canonical ensemble. The form and characteristics of 
this chemical potential-like piece are discussed in some detail. We argue that the resulting equations of motion derived from the action could be interpreted as the
Gross-Pitaevskii equation describing a graviton Bose-Einstein condensate trapped by the black hole 
gravitational field.
After this, we proceed to expand the ensuring equations 
of motion up to second order around the classical Schwarzschild metric so that some non-linear terms in the metric fluctuation are kept. 
Next we search for solutions and, modulo some very plausible assumptions, we find out that the condensate vanishes outside the horizon but is non-zero in its interior. Inspired by a linearized 
approximation around the horizon we are able to find an
exact solution for the mean-field wave function describing the graviton Bose-Einstein condensate
in the black hole interior. After this, we can rederive some of the relations involving the number 
of gravitons $N$ and the black hole characteristics along the lines suggested by Dvali and G\'omez.

\end{abstract}
\vfill
\noindent
May 2017

\noindent
ICCUB-16-032

\newpage

\section{Introduction}
In an interesting saga of papers, Dvali, G\'omez and others\cite{DG1,DG2}
have put forward an intriguing suggestion: black holes (BH) could perhaps be understood as
Bose-Einstein condensates of gravitons. If correct, this would reveal a deep quantum
nature of such fascinating objects and could lead to an alternative understanding
of some of the most striking features of BH; for instance, Hawking radiation \cite{Radiation} could
be understood as being due to leakage from the condensate. Besides, 
this picture brings new ideas about the Bekenstein entropy \cite{Entropy}, the absence of 
hair \cite{NoHair}, as well as the quantum nature of information storage and the possible 
information loss in BHs \cite{InformationLoss}.

The main point of these works is that the physics of BH can be understood in this picture
in terms of a single number $N$, the number of (off-shell) gravitons contained in the Bose-Einstein condensate (BEC).
These condensed gravitons have a wavelength $\lambda\sim \sqrt N L_P$, $L_P$ being the Planck length; they have 
a characteristic interaction strength $\alpha_g\sim 1/N$ and the leakage leads to a Hawking temperature 
of order $T_H \sim 1/\sqrt N L_P$, equal to
the inverse of $\lambda$. The mass 
of the BH is $M\sim \sqrt N M_P$ and its Schwarzschild radius therefore is given by
$r_s\sim \sqrt N  L_P$, thus agreeing with the Compton wavelength of the quantum  gravitons $\lambda$, in
accordance with the uncertainty principle that dictates $\lambda \simeq r_s$ in the ground state of
the quantum system. Therefore, up to various factors of the Planck mass everything is governed by $N$, the
number of intervening gravitons. Modulo some assumptions, all these results stem 
from the basic relation (unless otherwise stated we work in units where $\hbar=1$)
\begin{equation}
	r_g= \frac{N}{M}\end{equation}
that relates the number of gravitons $N$, the mass of the gravitating object $M$ and its gravitational
radius $r_g$. For a Schwarzschild BH $r_g=r_s$.

We found these results intriguing and we set up to try and understand them in a different language and, 
if possible, attempt to be more quantitative. On the process we have separated slightly from the
original line of thought of the authors. The original approach of \cite{DG1,DG2} is not geometric at all. There
is no mention of horizon or metric. Here we will adopt a more conservative approach. We will assume
the pre-existence of a classical gravitational field created by an unspecified source that generates
the Schwarzschild metric. As it is known, nothing can classically escape from a BH so if we wish to interpret
this in potential terms (which of course is not correct but serves us for the purpose of creating a picture
of the phenomenon) it would correspond to a confining potential. On and above this classical potential one
can envisage a number of quantum fields being trapped. For sure there is a gravity quantum field present
inside the horizon; hence gravitons.  Other quanta may get trapped by the BH potential as well, but a
possible graviton BEC seems particularly challenging to treat and this is the purpose of this work. 

Continuing with our semi-classical analogy, these trapped gravitons as well as other quanta present
have had a long time to thermalize in a (eternal) Schwarzschild BH and it is therefore natural 
to expect that after cooling they can form a Bose-Einstein condensate.
Of course these `gravitons' are in no way freely propagating transverse gravitons. 
They are necessarily off-shell ($q^2\neq 0$) and have some sort of effective mass. 
It may help to get a picture of the phenomenon to think of them as quasiparticles.

It is well known in BEC theory that Bose–Einstein condensation of a spatially homogeneous gas with attractive 
interactions is precluded by a conventional phase transition into either a liquid or solid. When confined to a 
trap; however, such a condensate can form, if its occupation number is low. Repulsive forces act to stabilize 
the condensate against collapse. Knowing this, one would immediately think that gravitons do not have 
repulsive interactions, at least naively, and that therefore a BEC is impossible to sustain, particularly as we expect $N$
to be very large. To this objection one could reply in a twofold way. First,
it is up to the equations to determine whether such a condensate is possible or not (we will see below
that indeed it is, at least in the case of vanishing angular momentum). Secondly one might also answer that,
in theories of emergent gravity, the ultimate nature of gravitons may be some type of fermionic degrees 
of freedom (such as e.g. in the model suggested in \cite{AEP}). Then, repulsion is assured at some scale
and fundamental collapse prevented.

We will try to identify and propose a  consistent set of equations describing a BEC constructed on top of the
classical field created by a BH\footnote{A classical field is not the same as a quantum condensate, although
the latter may trigger the former.}. We will succeed and we will see that remarkably enough the characteristics 
of the resulting BEC are uniquely
described in terms of the Schwarzschild radius of the BH and the value of a dimensionless parameter,
interpreted as a chemical potential. 
A condensate appears not only to be possible but actually intimately related to the classical field that sustains it and determines its characteristics. 

It would therefore be tempting to go one step beyond and reverse the order of the 
logical implication and eventually  attempt to derive the classical field as a sort of 
mean field potential \`a la Hartree-Fock. However we will stop short of doing so here. In any case, 
even without embarking in the discussion just mentioned, we are aware 
of the speculative nature of the present study.

\section{Building up a condensate over a Schwarzschild background}
In what follows we shall adhere to the following notational conventions: the Einstein tensor
$G_{\mu\nu}= R_{\mu\nu}-\frac12 g_{\mu\nu} R$ is constructed with the metric $g_{\mu\nu}$ in the usual way. 
We will denote by $\tilde g_{\mu\nu}$ the background metric that in 
our case will invariably be the Schwarzschild metric. Perturbations above this background metric will be denoted 
by $h_{\mu\nu}$, so $g_{\mu\nu}= \tilde g_{\mu\nu} + h_{\mu\nu}$. We will use the Minkowskian metric convention 
$\eta_{\mu\nu}=\rm{diag} (-1,1,1,1)$.

To initiate our program we should, first of all, identify an equation (or set of equations) that could provide a 
suitable description of a BEC in the present context. In other words, we have to find the appropriate generalization of 
the Gross-Pitaevskii equation. The graviton condensate has necessarily to be described by a tensor field that 
within our philosophy has to be connected necessarily with a perturbation of the classical metric. 

In order to keep things as simple as possible we will attempt to describe only condensates with quantum states having $l=0$. 
This will translate to spherically symmetric perturbations of the gravity field only
\be\label{Metric}
	ds^2=\left[-\left(1-r_s/r\right)+h_{tt}\right]\,dt^2+\left[\left(1-r_s/r\right)^{-1}+h_{rr}\right]\,dr^2+r^2d\Omega^2\,.\ee
The Einstein tensor derived from the previous metric will be expanded up to second order in $h_{\mu\nu}$ to retain the leading 
non-linearities (self-interactions of the desired condensate). 

It is well known that Birkhoff’s theorem \cite{ReggeWheeler,Birkoff} guarantees the uniqueness of the solution of Einstein's equations in vacuum with the properties of having spherical symmetry  and being static. 
As every dimensionful quantity can be expressed as a function of the Schwarzschild radius $r_s=2GM$, the difference 
between a given Schwarzschild metric and any perturbed solution built upon it must necessarily correspond to a change 
in the mass $M$ to $M+\delta M$. 
The fact that spherical symmetric perturbations are related to shifts in mass, indicates a correspondence between 
any perturbation $h_{\mu\nu}$ (i.e. each `graviton') and a certain amount of energy that is reflected in a change 
of the BH mass.

\subsection{Einstein equations as Gross-Pitaevskii equations}
The familiar Gross-Pitaevskii equation \cite{GP} employed to describe Bose-Einstein condensates is 
a non-linear Schr\"odinger equation; i.e. an equation of motion that contains self-interactions (hence the non-linearity),
a confining potential for the atoms or particles constituting the condensate, and a chemical potential, 
the conjugate thermodynamic variable of the number of particles or atoms contained in the condensate.

Among all these ingredients, the Einstein equations perturbed around a BH metric contain most of them. 
They are already non-linear and while there is no an explicit confining potential (as befits a relativistic theory) they do
confine particles, at least classically, because if the selected background corresponds to a Schwarzschild BH, 
the strong gravitational field classically traps particles inside the horizon. In addition, they are the only known 
consistent equations describing a rank-two tensor. Finally, the background (i.e. the BH metric) 
is a solution of Einstein equations; therefore they are necessarily part of the answer. However, there is one ingredient 
still missing, namely the equivalent of the
chemical potential. Therefore we have to extend the formulation of perturbations around a classical BH solution to the 
grand-canonical ensemble by adding to the appropriate action a chemical potential term.

As it is well known since the early days of quantum field theory \cite{PW} there are no conserved currents or 
continuity equations for fundamental fields that 
are chargeless (such as a real Klein-Gordon field). Therefore there is no way of defining a number operator
for freely propagating gravitons or photons. Nonetheless, in the picture of \cite{DG1} and \cite{DG2} the situation is different. {\em If} a BEC is present with
`gravitons' acquiring all the same momenta $\sim 1/r_s$ and being weakly interacting for a macroscopic BH 
(recall $\alpha_g\sim 1/N$) the total energy stored in the condensate should be $\sim N/r_s$ and this quantity
would be a conserved one. Therefore, lest $r_s$ change, $N$ would be conserved. The previous reasoning 
shows very clearly that the `gravitons' contemplated in the present scenario, if realized,  have nothing to do
with freely propagating gravitons.

The energy contained in a given volume occupied by a non-interacting scalar field is 
\begin{equation}
	E= -\frac12 \int dV\phi\nabla^2\phi\,;\end{equation} 
by analogy, in the present case
\begin{equation}
	E=\frac12 \int dV \varepsilon^2 \hat h_{\alpha\beta}\hat h^{\alpha\beta}= \int dV \varepsilon \,\rho_{\hat h}\, ,\end{equation}
where we assume that the energy per graviton $\varepsilon$ is constant and approximately given by $1/r_s$, and
\begin{equation}\label{Density}
	\rho_{\hat h}\equiv \frac12\hat h_{\alpha\beta}\frac{1}{\lambda} \hat h^{\alpha\beta}\,,\end{equation}
with $\hat h_{\alpha\beta}= M_P h_{\alpha\beta}$.
While there is no formally conserved current, the above quantity can be interpreted as a `graviton' number density, and
the integral of the graviton density (\ref{Density}) in the interior of the BH has to be interpreted 
as the number of constituents of the BEC.

The above considerations can now be phrased in a Lagrangian language.
The chemical potential term in the action should be related to the graviton density of the 
condensate inside a differential volume element $dV$. In order to respect the basic symmetry of General Relativity (GR), 
the simplest form of introducing such a term is by means of
\begin{equation}\label{ActionCPTerm}
	\Delta S_{chem. pot.}
	= -\frac12 \int d^4 x\sqrt{-g}\,\mu \,\hat h_{\alpha\beta}\hat h^{\alpha\beta}\,.\end{equation} 
This term does of course resemble a mass term for the spin-2 excitation and indeed it is some sort of effective
mass in practice as the `gravitons' in the BEC are quasi-non-interacting. However we will see that GR eventually requires
for the quantity $\mu$ to actually be position dependent, i.e. $\mu=\mu(r)$, and transform as a scalar.

Given the resemblance of the term that we interpret as chemical potential to a mass term, it is legitimate to ask why not use 
the Fierz-Pauli form of the mass term, that is known to be able to provide ghost-free 
propagation of gravitons (see e.g. \cite{KH} for a 
detailed discussion of massive gravity). The answer is simple: our `gravitons' correspond to time-independent 
solutions; they do not 
propagate and the issue of ghosts is totally irrelevant in the present discussion. We do not advocate adding 
(\ref{ActionCPTerm}) as a fundamental ingredient of gravity, but only as a means of describing in an effective 
and thermodynamical way the formation of a BEC condensate ---exactly what the Gross-Pitaevskii equation is meant to implement.
From a more pragmatic point of view, modifications of (\ref{ActionCPTerm}) in the sense of making this term look like a Fierz-Pauli
mass term do not really change the results at the qualitative level, but we emphasize that it is not really necessary to worry
about the ghost issue at this point.

Although the additional term may look odd,
it is invariant under the gauge group of diffeomorphisms. Part of this statement is shown in \cite{Weinberg}: under 
an infinitesimal displacement in the coordinates of the form $\delta_{\scriptscriptstyle D} x^\mu=\xi^\mu$ the full metric 
changes as $\delta g_{\mu\nu}=\xi_{\mu;\nu}+\xi_{\nu;\mu}$. The same gauge transformation rules the background metric. 
As the perturbation is defined by $h_{\mu\nu}= g_{\mu\nu}-\tilde g_{\mu\nu}$, under the same perturbation of the 
coordinate system, the same rule of covariant transformation is obtained.  Together with the fact that the 
chemical potential $\mu$ behaves as a scalar under a general coordinate transformation, 
ensure automatically the general covariance of the theory. However, while the formalism is diff invariant, it is 
{\em not} background independent. The separation $g_{\mu\nu}=\tilde g_{\mu\nu}+h_{\mu\nu}$  leads to an action that depends on 
the choice of the background metric, which is
the one of the BH. Likewise $\mu(r)$ also will depend on the background metric. We will not postulate any particular dependence of $\mu$ on $r$. In fact, at this point it would be conceivable that the
only consistent solution implies $\mu=0$. This will not be the case however and $\mu(r)$ will be determined in the subsequent
from the requirement of self-consistency of the proposal. Therefore an appropriate action for the field $h_{\alpha\beta}$ is
\begin{equation}\label{FullAction}
	S(h)= {M_P}^2\int d^4 x \sqrt{-g}\,R(g)- \tfrac12 {M_P}^2 \int d^4 x\sqrt{-g} \,\mu\, h_{\alpha\beta}h^{\alpha\beta}\,.\end{equation}
Indices are raised and lowered using the full metric $g_{\mu\nu}=\tilde g_{\mu\nu}+h_{\mu\nu}$. A completely covariant expansion in powers of $h_{\alpha\beta}$ can be performed up to the desired order of accuracy. 
By construction these equations will be non-linear, but we will keep only the leading non-linearities to maintain 
the formalism simple and analytically tractable. It is worth noting again that the action (\ref{FullAction}) 
is as shown reparametrization invariant, but it is not background independent. The fact that fluctuations take place above a 
BH background --in whatever coordinates one chooses to describe it-- does matter.

The action principle yields the two equations of motion for the field $h_{\alpha\beta}={\rm diag}(h_{tt},h_{tt},0,0)$, namely
\begin{equation}\label{EEq}
	G_\alpha\,^\beta(\tilde g+h)= \mu \bigl(h_\alpha{}^\beta-h_{\alpha\sigma}h^{\sigma\beta}+\tfrac14\,h^2\,\delta_\alpha^\beta\bigr)\,,\end{equation}
where $\delta_\alpha^\beta$ is the Kronecker delta and $h^2=h_{\alpha\beta}h^{\alpha\beta}$. 
It is important to keep in mind the following: we are working here in the grand canonical 
ensemble; this implies that the magnitude $\mu$ is an external field and does not vary in the action. 
In particular, for these equations of motion $\mu$ is independent of $h_{\alpha\beta}$, so $\delta \mu/\delta h_{\alpha\beta}=0$. 
This should be the way of introducing the chemical potential. Otherwise, were $\mu$ not an external field, 
it would be necessary to take it into account when performing variations to derive the equations of motion. 
An equation of motion would be obtained for $\mu$ and this equation would nullify automatically the perturbation 
as well as the chemical potential itself, i.e. $h_{\alpha\beta}=0$ and $\mu=0$. Therefore, the external field $\mu$ 
is introduced in the theory as some kind of Lagrange multiplier. As Lagrange multipliers have constraint equations, 
the chemical potential of the theory may not be arbitrary at all. 
It must satisfy binding conditions with $h_{\alpha\beta}$. The main difference is that these constraints are implicit in a GR theory. 
The restriction of $\mu$ is through the general covariance conditions for the action. In other words, the 
diffeomorphisms covariance implies the covariant conservation of the Einstein tensor, and this in itself entails the same 
for the chemical potential term
\begin{equation}\label{DiffEqChemPot}
	\nabla_\beta G_\alpha{}^\beta=0\quad\Longrightarrow\quad
	\Bigl[\mu\Bigl(h_\alpha{}^\beta-h_{\alpha\sigma}h^{\sigma\beta}+\tfrac14\,h^2\,\delta_\alpha^\beta\Bigr)\Bigr]_{;\beta}=0\,.\end{equation}
The covariant derivative is defined using the full metric $g_{\alpha\beta}=\tilde g_{\alpha\beta}+h_{\alpha\beta}$. This differential equation of motion is valid up to every order in perturbation theory.

Hereinafter, we will proceed to find acceptable solutions of these equations and interpret them. We will 
separate the problem into two regions: outside and inside the BH horizon. We certainly expect that the 
graviton condensate will disappear quickly in the outside region; imposing this as a boundary condition 
for $r\gg r_s$ we will see that in fact the condensate is identically zero on this side of the horizon. 
On the contrary, a unique non-trivial solution will be found in the interior of the BH.

\section{Outside the horizon}\label{OutsideHorizon}
In this section it will be shown that even after the inclusion of $\mu$ there is not other normalizable solution 
outside the black hole horizon than the trivial solution for the perturbation. If we place ourselves far away from the BH, 
in the $r\to\infty$ limit, one is able to keep only dominant terms in the Einstein equations. 
In this perturbative treatment the chemical potential, far away from the condensate, is expected to behave as a 
perturbative small parameter. Therefore, on the LHS side of the equations of motion (\ref{EEq}), possible terms of 
the form $\mu h^2$ will be considered as $O(h^3)$ and neglected.
We will make use of another equation that, although it does not contribute with new information, will make 
explicit how the components of the perturbation behaves between each other; namely, any of the two angular components of the 
Einstein tensor $G_{\theta}\,^{\theta}=G_{\phi}\,^{\phi}=0$. In this limit, as we want a vanishing perturbation at infinity, 
the following ansatz is imposed in the faraway region:
\begin{equation}
	h_{tt}=\frac{A}{r^a}\,;\hspace{30pt} h_{rr}=\frac{B}{r^b}\,;\hspace{30pt} \mu=\frac{C}{r^c}\,;\hspace{30pt} \hbox{with }\,a,\,b,\,c >0 \,.\end{equation}

Before proceeding, an additional consideration is needed:
since for us $h$ represents a localized BEC, in analogy with the wave function of a confined particle, the perturbation must be a square-integrable function. This is
\begin{equation}\label{h2Integral}
	\int_{0}^{\infty}d^3x \sqrt{-g}\,{h}^2
	\equiv\int_{0}^{\infty}d^3x \sqrt{-g}\, \left({h_{tt}}^2\,g^{tt\,2} +{h_{rr}}^2\,g^{rr\,2}\right)<\infty\,.
\end{equation}
There are two reasons that lead us to the previous requirement. Let us review first why one has to request square-integrability.
Note that the solutions of (\ref{EEq}) should not be understood as a quantum field, but rather as solutions of 
the Gross-Pitaevskii equation. Here we adopt Bogoliubov theory \cite{NNB} and its interpretation of the GP wave function. 
This is the commonly accepted interpretation and essentially it boils down to the fact that in the large occupation 
limit ($N\to \infty $) the creation and annihilation operators of the ground state can be approximately treated 
as commuting c-numbers, and hence the many body quantum problem is described by a classical function called the 
`macroscopic wave function' or simply the `order parameter'. Because the modulus square of this quantity is 
proportional to ${a_0}^\dagger\, a_0$, it is therefore proportional to $N$. Hence the order parameter 
itself is proportional to $\sqrt{N/V}$. 
If the potential is not uniform, $N$ may of course depend on the coordinates. 
From this the need to require square-integrability follows.
This interpretation is in the present case also supported by the dimensionality of $\rho_{\hat h}$.

Why then this precise condition? The leading role of the geometry has been to provide, in an indirect manner, 
a ``confining potential" that traps gravitons inside 
the horizon. Therefore, from this point of view when giving a physical interpretation to the chemical potential 
(see section 4) one implicitly assumes that the geometry of the spacetime is flat and that geometry acts via
the external potential and the chemical one. In the present situation, it turns out that 
$\sqrt{g} \propto r^2\sin\theta$ coincides with the one corresponding to a 3 dimensional spatial 
flat metric in spherical coordinates; therefore $d^3x\sqrt{g}$ is reparametrization invariant in 3 dimensions. 
This interpretation is equivalent to defining a wave function normalized by the temporal component of the metric 
tensor, $\psi=\sqrt[4]{g_{tt}}\,\,h$. Then, any magnitude computed by means of an integration over a 3-spatial volume $d^3V=d^3x\,\sqrt{-\gamma}$ is effectively written in a fully covariant way as $d^3V\abs{\psi}^2=d^3x\sqrt{-\gamma}\,\sqrt{g_{tt}}\,h^2=d^3x\sqrt{g}\,h^2$, since the 4-dimensional metric can be decomposed as $g_{\mu\nu}=g_{tt}\gamma_{ij}$, with $i,j=r,\theta,\phi$ if the 
metric is diagonal. Therefore, the volume element is coordinate dependent as well as the  square modulus of the macroscopic 
wave function $\psi$, but the combination of both is not; and we end up finally with the corresponding volume element to the previously 
discussed flat spacetime.

Whatever the case, in this perturbative limit the components of the Schwarzschild metric accomplish $\tilde g^{tt\,2},\,\tilde g^{rr\,2}\xrightarrow{r\rightarrow\infty} 1$; then the integral in Eq. (\ref{h2Integral}) should fulfill
\begin{equation}
	4\,\pi\int_{0}^{\infty}dr\,r^2\,\left[{h_{tt}}^2+{h_{rr}}^2+O(h^3)\right] <\int_{0}^{\infty} \frac{dr}{r} =\infty \quad(\hbox{Logarithmic divergence}) \,.\end{equation}
For this to happen, if a power law ansatz at infinity is imposed, the exponents must obey $a,\,b>3/2$. 
Then the equations corresponding to $G_t{}^t$, $G_r{}^r$ and $G_\theta{}^\theta$ respectively become
\begin{equation}\label{EEqInfinity}\begin{split}
	\left(1-b\right)\,\frac{B}{r^{b+2}}-\left(1-2\,b\right)\,\frac{B^2}{r^{2b+2}}&=\frac{A\,C}{r^{a+c}}
	\\[4ex]
	\left(\frac{r_s}{r}+a\right)\frac{A}{r^{a+2}}-\frac{B}{r^{b+2}}+\left(\frac{r_s}{r}+a\right)\frac{A^2}{r^{2a+2}}-\left(\frac{r_s}{r}+a\right)\frac{A\,B}{r^{a+b+2}}+{\frac{B^2}{r^{2b+2}}}&=\frac{B\,C}{r^{b+c}} 
	\\[4ex]
	\left(\frac{r_s}{r}+a^2\right)\frac{2\,A}{r^{a+2}}+\left(\frac{r_s}{r}-b\right)\frac{2\,B}{r^{b+2}}+\left(\frac{2r_s}{r}+3\,a^2\right)\frac{A^2}{r^{2a+2}}-\left({\frac{r_s}{r}}-b\right)\frac{4\,B^2}{r^{2b+2}}&
	\\
	-\left[\left(2+b\right)\frac{r_s}{r}+2\,a^2+a\,b\right]\frac{A\,B}{r^{a+b+2}} &=\hspace{5pt}0 \,.\end{split}\end{equation}
The requirement of the solution being  square-integrable exclude that $a$ and $b$ could  be zero. 
This implies automatically the neglection of the terms proportional to the Schwarzschild radius. 
Even more, with the previous considerations, (\ref{EEqInfinity}) reduce to the following asymptotic identities
\begin{equation}\label{Gtt Schw powerLaw ansatz O(1)}
	\left(1-b\right)\frac{B}{r^{b+2}}=\frac{A\,C}{r^{a+\rho}} \,;\hspace{40pt}
	\frac{A\,a}{r^{a+2}} -\frac{B}{r^{b+2}}=\frac{B\,C}{r^{b+c}}\,;\end{equation}

\begin{equation}\label{Gthth Schw powerLaw ansatz O(1)}
	\frac{2\,A\,a^2}{r^{a+2}} -\frac{2\,B\,b}{r^{b+2}}=0\,.\end{equation}
From the angular equation (\ref{Gthth Schw powerLaw ansatz O(1)}), it is mandatory for both terms to contribute at infinity. 
If this is not the case, this would nullify $A$ (or $B$), and then, via both equations 
in (\ref{Gtt Schw powerLaw ansatz O(1)}), fix $B=0$ (or $A=0$) and $C=0$; i.e. $h_{\alpha\beta}=\mu=0$. 
The competition of both terms is possible if and only if $a=b$. 
This condition modifies the angular equation (\ref{Gthth Schw powerLaw ansatz O(1)}) as
\begin{equation}
	\left(A\,a-B\right)\frac{2\,a}{r^{a+2}}=0 
	\qquad\Longrightarrow\qquad
	A\,a=B\,.\end{equation}
Then, from the second equation in (\ref{Gtt Schw powerLaw ansatz O(1)}) the following relationship can be read
\begin{equation}
	\left(A\,a-B\right)\frac{1}{r^{a+2}}=0=\frac{B\,C}{r^{a+c}}\,.\end{equation}
This automatically leads to a null chemical potential as the only possible solution for this region with the proposed ansatz.

We have also explored the possibility of exponentially vanishing chemical 
potential when $r\to\infty$ with analogous conclusions. 
If we change the ansatz and impose an exponential decreasing solution for the perturbations at infinity,
\begin{equation}
	h_{tt}=A\,e^{a\,r} \,;\hspace{30pt} h_{rr}=B\,e^{b\,r} \,;\hspace{30pt} \mu=C\,e^{c\,r} \hspace{20pt} \hbox{with } a,\,b,\,c<0\,,\end{equation}
the linear order dominates in front of the higher ones. As decreasing solutions are expected, the parameters $a$, $b$ and $c$ 
must be not null. If one use this ansatz in the Einstein equations partially evaluated at infinity, one finds that in 
the angular equation there is no way for both leading terms to compete between each other
\begin{equation}
	G_\theta{}^\theta:-2\,A\,a^2\,e^{a\,r}-2\,B\,b\,\frac{e^{b\,r}}{r}=0\,.\end{equation}
In conclusion, depending on the fact whether if $a$ is bigger or not than $\beta$, $A=0$ or $B=0$. For the first case, if $A=0$, the leading terms of the temporal equation
\begin{equation}
	G_t{}^t:\,B\,b\,\frac{e^{b\,r}}{r}=A\,C\,e^{\left(a+c\right)\,r}\end{equation}
fixes $B=0$. Being both, $A$ and $B$ null, the chemical potential disappears from the theory in this region. 
For the second case, if $B=0$, the same equation $G_t{}^t$ nullifies $A$ (as no null solutions for $C$ are expected). 
No perturbation makes the chemical potential senseless. To sum up, this ansatz implies as well that the only solution 
at infinity is a null perturbation, $h_{\alpha\beta}=0$, and the disappearance of the chemical potential from the theory, 
$\mu=0$. Likewise it can be seen that a much faster Gaussian decay is also excluded with analogous calculations.

Before concluding this section we will give additionally a numerical argument that confirms the vanishing 
of the condensate and the chemical potential in the outer region when the perturbations are null at infinity. 
The following change of variables allows rewriting the relevant equations (\ref{EEq}) in terms of dimensionless quantities
\begin{equation}
	z=\frac{r_s}{r}\,; \qquad X(z)=\mu(r)\, r^2\,.\end{equation}
This quantities are
universal as $r_s$ and $\mu$ are the only physical parameters. This redefinition maps the exterior part into a 
compact interval: $r=\infty$ is transformed into $z=0$ and $r=r_s$ into $z=1$. It is well known that Schwarzschild 
space-time is asymptotically flat; therefore, at infinity ($z=0$) it is expected for 
the perturbation to vanish ($h_{\alpha\beta}=0$). Nevertheless equations are difficult to deal with because 
they are singular at $z=0$, i.e. it is not possible to isolate the higher order derivatives.

Let us see how we can proceed. As mentioned before, we have to settle 
for a point  $z\sim0$ (but $z\neq 0$) to set up boundary conditions  for the numerical integration procedure to start. 
Likewise setting $h_{rr}=0$ at any arbitrary value of $z$ outside the event horizon, immediately triggers divergences at the 
first step of the routine. Therefore we take a ``small" initial value for the perturbation in a point near 
infinity ($z\sim0$), and then decrease this initial value; this is $h_{\alpha\beta}(r\sim \infty)\rightarrow 0$. 
\begin{figure}[!b]
	\begin{center}
		\vspace{-15pt}
		\includegraphics[scale=0.5]{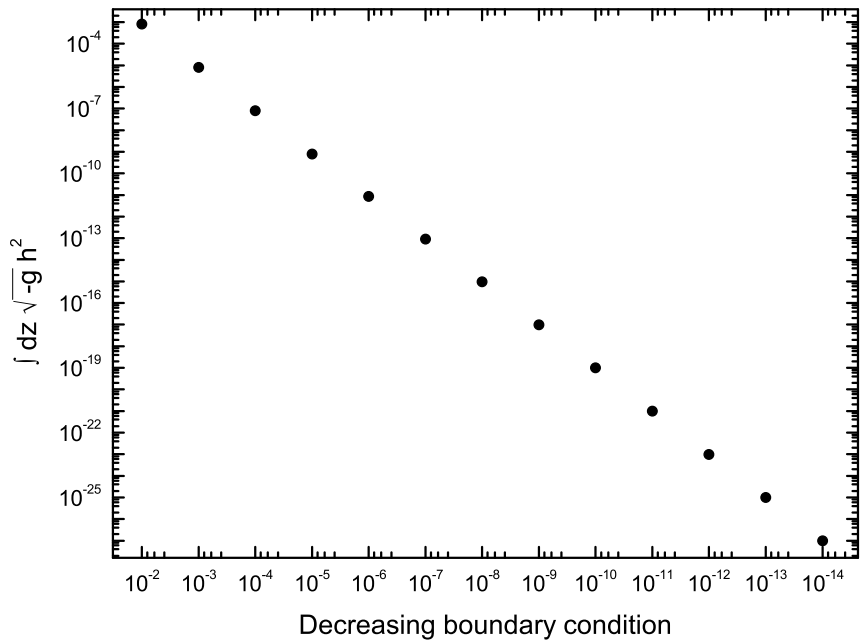}
		\vspace{20pt}
		\captionsetup{width=0.85\textwidth}
		\caption{Each point represents a magnitude proportional to the integral of $h^2$ outside the event horizon, computed for a particular initial condition for the perturbation placed near infinity, $h_{\mu\nu}(\infty)$. The graph is presented in a log--log scale. A linear fit gives a slope that asymptotically approach $2$ as the boundary condition is nullified; this implies that the integral vanishes quadratically. The closer the limit to an asymptotically flat space-time is (i.e. decreasing the initial condition near infinity), the smaller this integral becomes.}\label{NOut}\end{center}\end{figure}
With the set of solutions for the components of $h_{\alpha\beta}$ obtained 
for each of the initial condition imposed, it is possible to compute the quantity $\int dV \, h^2$
defined in Eq. (\ref{h2Integral}); the integral extends over the exterior of the BH (i.e. $z$ runs between 0 and 1). 
In Figure \ref{NOut} the behaviour of the integral of $h^2$ outside the BH is presented when the initial condition for the integration approaches zero at a fixed point far away from the condensate (near infinity or $z\sim0$). The numerical analysis results very stable and the initial condition can be reduced as many orders of magnitude as desired. The graph is presented in a log--log scale, hence the linear behaviour. 
The points are fitted by a linear function with a slope approaching $2$ as more points with lower initial condition are added. This implies that the integral converges quadratically to zero as the initial condition for $h_{\alpha\beta}$ nullifies. 
This seems to confirm ---also numerically--- that there is no condensate, i.e. $h_{\alpha\beta}=0$ and no dimensionless chemical potential $X=0$ in the outer region of the BH.

\section{Inside the horizon}
In order to study the behaviour of our equations when the horizon is crossed, it is convenient to keep working in 
the $z=r_s/r$ and $X=\mu\,r^2$ variables introduced in the previous section and redefine the perturbations as
\begin{equation}\label{RedefinitionPert}
	h_{tt}(r)=\left(1-z\right)\gamma_{tt}(z)\,; \hspace{40pt} h_{rr}(r)=\left(1-z\right)\gamma_{rr}(z)\,.\end{equation}
With the new definitions the Einstein equations become more compact. At the same order as before, the temporal, radial 
and angular components of the Einstein tensor read as
\begin{equation}\label{G with inverse metric}\begin{split}
	(2z+1)(z-1)^2\gamma_{rr} +z(z-1)^3{\gamma_{rr}}' -(4z+1)(z-1)^4{\gamma_{rr}}^2 -2z(z-1)^5\gamma_{rr} \,{\gamma_{rr}}'
	\\[3ex]
	-(z-1)^2\gamma_{rr} -z(z-1){\gamma_{tt}}' -z(z-1)\gamma_{tt}\,{\gamma_{tt}}' +z(z-1)^3\gamma_{rr}\,{\gamma_{tt}}' +(z-1)^4{\gamma_{rr}}^2 
	\\[3ex]
	2z(z-2)(z-1)\gamma_{rr} +z(z-2)(z-1)^2{\gamma_{rr}}' -z(5z-2){\gamma_{tt}}' -2z^2(z-1){\gamma_{tt}}'' \\ -2z(z-2)(z-1)^4\gamma_{rr}{\gamma_{rr}}' +z(7z-2)(z-1)^2\gamma_{rr}{\gamma_{tt}}'
	-z(5z-2)\gamma_{tt}\,{\gamma_{tt}}' \\ +z^2(z-1)^3{\gamma_{rr}}'\,{\gamma_{tt}}' -z^2(z-1){{\gamma_{tt}}'}^2 +2z^2(z-1)^3\gamma_{rr}\,{\gamma_{tt}}'' \\ -2z^2(z-1)\gamma_{tt}\,{\gamma_{tt}}''\,.\end{split}\end{equation}

\subsection{Linearization near the event horizon}
In order to get a feeling for possible solutions to these equations we consider their linearized
approximation. Only terms linear in $\gamma_{tt}$, $\gamma_{rr}$ and their derivatives are kept in the region $z\to 1$. 
No a priori assumption for the dimensionless chemical potential $X$ is made, therefore $X\gamma_{\alpha\beta}$ is a priori 
considered as a linear contribution in the perturbation. That is
\begin{equation}\begin{split}
	3(z-1)^2\gamma_{rr}+(z-1)^3{\gamma_{rr}}'=X \gamma_{tt} \\[3ex]
	(z-1)^2\gamma_{rr}+(z-1){\gamma_{tt}}'= -X(z-1)^2\gamma_{rr} \\[3ex]
	2(z-1)\gamma_{rr}+(z-1)^2{\gamma_{rr}}'+3{\gamma_{tt}}'+2(z-1){\gamma_{tt}}''=0 \,.\end{split}\end{equation}
Let us now make the following ansatz for the new perturbations
\begin{equation}
	\gamma_{tt}\sim A (z-1)^a \,; \qquad
	\gamma_{rr}\sim B (z-1)^b\,,\end{equation}
while we leave the chemical potential $X=X(z)$ as a free function. The three equations take respectively the following form
\begin{equation}\begin{split}
	B(3+b)(z-1)^{b+2}=X A(z-1)^a
	\\[2ex]
	B(z-1)^{b+2}+Aa(z-1)^{a}=-X B(z-1)^{b+2}
	\\[2ex]
	B(2+b)(z-1)^{b+1}+Aa(1+2a)(z-1)^{a-1}=0\,.\end{split}\end{equation}
It is worth noting that if $a=b+2$, all terms contribute as we get closer 
to the event horizon and $X$ behaves as a constant. In this situation, we obtain three equations for the coefficients 
\begin{equation}
	B(1+a)=X A\,;\qquad B+a A= - X B \,;\qquad a[B+A(1+2a)]=0\,.\end{equation}
The system of equations is algebraic for the variable $X$ (that as we have seen should behave as a constant as $z\to 1$); 
therefore, it is possible to eliminate $X$ by combining the temporal and radial equation. In the present ansatz this leads to
\begin{equation}
	B^2(1+a)+AB+A^2a=0\,.\end{equation}
Together with the angular equation, this determines the solutions up to a single constant.
There are two possible solutions for this system
\begin{itemize}
\item[i)]	\hfill \makebox[0pt][r]{%
\begin{minipage}[b]{\textwidth}\begin{equation}\label{behaviourw=1}
	a=0 \hspace{25pt} b=-2 \hspace{20pt} A=-B \hspace{20pt}\Longrightarrow\hspace{20pt} \gamma_{tt}=A \hspace{35pt} \gamma_{rr}=-\frac{A}{(z-1)^2}\end{equation}\end{minipage}}
		
\item[ii)]	\hfill \makebox[0pt][r]{%
\begin{minipage}[b]{\textwidth}\begin{equation} 
	a=-1  \hspace{20pt} b=-3  \hspace{20pt} A=B  \hspace{20pt}\Longrightarrow\hspace{20pt} \gamma_{tt}=\frac{A}{(z-1)} \hspace{20pt} \gamma_{rr}=\frac{A}{(z-1)^3}\,.\end{equation}\end{minipage}}\end{itemize}
In any case, at least one of the perturbations is divergent over the event horizon. Nonetheless, the first solution 
appears to be integrable, while this is not the case for the second one. In case i) $X=-1$ while if ii) is taken as 
solution $X=0$ (i.e. no chemical potential at all).\footnote{Solution ii) represents however a volume-preserving fluctuation
at the linear order.}

However, because these solutions do not vanish when $z\to 1$, justified doubts can be cast on the relevance of the 
linearized equations. Let us 
examine this point taking into consideration only the solution i), which is integrable according
to the considerations of the previous section.

In order to see if this solution is modified when non-linearities are switched on, we substitute it back 
in the $O(\gamma^2)$ equations system where self-interactions matters, and see if the solution survives or how 
would it get modified. In view of the foregoing, near $z\sim 1$ the full temporal, radial and angular equations read
\begin{equation}\begin{split}
	-3(z-1)^2\gamma_{rr} -(z-1)^3{\gamma_{rr}}' +5(z-1)^4{\gamma_{rr}}^2 +2(z-1)^5\gamma_{rr}\,{\gamma_{rr}}' \\
	=X\Bigl\{-\gamma_{tt}-2{\gamma_{tt}}^2+\tfrac14\left[{\gamma_{tt}}^2+\left(z-1\right)^4{\gamma_{rr}}^2\right]\Bigr\}\end{split}\end{equation}

\begin{equation}\begin{split}
	-(z-1)^2\,\gamma_{rr} -(z-1){\gamma_{tt}}' -(z-1)\gamma_{tt}\,{\gamma_{tt}}'
	+(z-1)^3\gamma_{rr}\,{\gamma_{tt}}' +(z-1)^4{\gamma_{rr}}^2 \\ =X\Bigl\{\left(z-1\right)^2\gamma_{rr}-2\left(z-1\right)^4{\gamma_{rr}}^2+\tfrac14\left[{\gamma_{tt}}^2+\left(z-1\right)^4{\gamma_{rr}}^2\right]\Bigr\}\gamma_{rr} \end{split}\end{equation}

\begin{equation}\begin{split}
	2(z-1)\gamma_{rr} +(z-1)^2{\gamma_{rr}}' +3{\gamma_{tt}}' +2(z-1){\gamma_{tt}}'' -4(z-1)^3{\gamma_{rr}}^2 -2\,(z-1)^4\gamma_{rr}\,{\gamma_{rr}}' \\
	+3\gamma_{tt}\,{\gamma_{tt}}' -5(z-1)^2\gamma_{rr}\,{\gamma_{tt}}' -(z-1)^3{\gamma_{rr}}'\,{\gamma_{tt}}' +(z-1){{\gamma_{tt}}'}^2 +2(z-1)\gamma_{tt}\,{\gamma_{tt}}'' \\ -2(z-1)^3\gamma_{rr}\,{\gamma_{tt}}''=0, \end{split}\end{equation}
respectively. Replacing the possible solution, $\gamma_{tt}=A$ ---this eliminates any derivative of $\gamma_{tt}$---
and  $\gamma_{rr}=-\frac{A}{(z-1)^2}$ into these equations, we obtain the following relations
\begin{equation}\begin{split}
	3\,A - 2\,A +5\,A^2 -4\,A^2 =(1+A)A =-X\,\left(A+\tfrac32A^2\right) \end{split}\end{equation}
\begin{equation}\begin{split}
	A+A^2 =(1+A)A =-X\,\left(A+\tfrac32A^2\right) \end{split}\end{equation}
\begin{equation}\begin{split}
	-\frac{2\,A}{(z-1)} +\frac{2\,A}{(z-1)} -\frac{4\,A^2}{(z-1)} +\frac{4\,A^2}{(z-1)} =0 \,.\end{split}\end{equation}
Therefore, quite surprisingly, the linear solution is still an exact solution of 
the non-linear quadratic differential equations. Thus, we conclude that 
\begin{equation}\label{Event Horizon}
\gamma_{tt}=A \hspace{100pt} \gamma_{rr}=-\frac{A}{(z-1)^2} \end{equation}
are solutions of the second order system of equations. 

In addition, this exercise gives an interesting result: $X\simeq -(1-A/2)$, where $A$ is so far arbitrary, also 
in sign. Note that at the linear level we got $X=-1$ and the fact that
the quadratic equation gives an $O(A)$ correction to this result is consistent as we are implicitly assuming
that $|h_{\alpha\beta}|\ll |\tilde g_{\alpha\beta}|$, i.e. $|A|\ll 1$. The constant $A$
itself is arbitrary and is not determined by the structure of the equations. Changes in $A$ appear 
as an overall factor in the solution. 

So far we have seen that the solution given in (\ref{Event Horizon}) satisfies not only the linearized
approximation but also the full quadratic equations. This conclusion is reinforced after
performing a numerical integration of the basic equations (\ref{EEq}) expanded up to $O(h_{\alpha\beta}{}^2)$.
We found that the numerical study reproduces the general features of the analytical study: $\gamma_{tt}$
as well as the function related to the dimensionless chemical potential $X$, turns out to be constant in the interior of the BH.
No other solutions are found.

The equations look simpler if expressed in terms of the functions $\gamma_{\alpha\beta}$ but to understand
what this solution means is better to undo the redefinition of the components of 
the wave function (i.e. $\gamma_{\alpha\beta}\rightarrow h_{\alpha\beta}$) by means of (\ref{RedefinitionPert}) for 
the solution (\ref{Event Horizon}). It is of interest to raise one index of the components of the perturbation 
with the inverse full metric $g_{\mu\nu}$. The function so obtained happens to be constant throughout the interior of the BH
\begin{equation}
	h_{\alpha\beta}g^{\alpha\beta}=h_t{}^t = h_r{}^r = {\rm constant}\,.\end{equation}
The solution for the perturbation $h_{\alpha\beta}$ turns out  to be proportional to the
corresponding metric element where each of the two belongs; i.e. $h_{tt}=h_t{}^t g_{tt}$
and $h_{rr}=h_r{}^r  g_{rr}$ with $h_\alpha{}^\alpha$ constant. The angular degrees of freedom remain unchanged, $h_\theta{}^\theta=h_\phi{}^\phi= 0$.

\subsection{Exact solution}
Inspired by the previous analysis, we reformulate our main equations in terms of the metric
fluctuation with mixed indices (one covariant, one contravariant). Let us write
\begin{equation}\label{CoContraPert}
	h_t{}^t=\varphi_t \,;\qquad h_r{}^r=\varphi_r \,; \qquad h_\theta\,^\theta =h_\phi\,^\phi=0\,.\end{equation}
The full metric would become
\begin{equation}\label{metricfull}
	g_{\mu\nu}={\rm diag}\Bigl(\,\frac{1}{1-\varphi_t}\tilde g_{tt}\,,\,\frac{1}{1-\varphi_r}\tilde g_{rr}\,,\, \tilde g_{\theta\theta}\,,\,\tilde g_{\phi\phi}\,\Bigr)\end{equation}
and seems to impose an upper limit\footnote{The upper bound for the wave function will become clear when computing the number $N$ of constituents of the condensate; such value for the wave function corresponds to the limit $N\rightarrow\infty$.} for the
constant values $\varphi_t$ and $\varphi_r$. The exact equations of motion for the theory simplify enormously and reduce to the following ones
\be\begin{aligned}\label{EEqhCte}
	&{G_t}^t =-\frac{\varphi_r}{r^2} =\mu\left[\varphi_t +\frac{1}{4}\left(-3\varphi_t^2+\varphi_r^2\right)\right]
	\\&{G_r}^r=-\frac{\varphi_r}{r^2}
	=\mu\left[\varphi_r +\frac14\left(\varphi_t^2-3\varphi_r^2\right)\right]\,.\end{aligned}\ee 
These Eqs. are valid also for the external solution; however, in the case of the external sector, we know that Minkowski metric must be recovered far from the sources. This condition implies that $\varphi_t=\varphi_r=0$ as explained in Section \ref{OutsideHorizon}.

Among the two possible solutions of the latter algebraical system of equations, only one is compatible with 
an acceptable limit for small perturbations, namely $\varphi_t=\varphi_r\equiv\varphi$ (we expect $0<\varphi<1$). 
All things considered, 
the two equations in (\ref{EEqhCte}) are linearly dependent, hence the resulting equation of motion for a 
constant $\varphi$ becomes
\begin{equation}\label{GPExact}
	-\frac{\varphi}{r^2}= \mu \varphi -\frac12 \mu \varphi^2\,.\end{equation}
Within our philosophy (\ref{EEq}) is understood as a Gross-Pitaevskii equation for a condensate wave 
function $h_\alpha{}^\alpha\equiv\varphi$. It is a non-linear Schr\"odinger-like equation that produces a 
unique solution for the condensate and the chemical potential described by Eq. (\ref{GPExact}). Because
the solution is constant, the `kinetic' term drops and one is left with a purely algebraic, 
mean-field-like, equation\footnote{One usually thinks of the GP equation as a non-linear Schr\"odinger equation, hence with 
second order derivatives. Actually this is not always so; for uniform gas of interacting atoms the GP equation is simply
$g\varphi^2 = \mu$, where $g$ is the interacting (repulsive) constant and $\mu$, needless to say, is the chemical potential.}.

The previous equation defines a dimensionless chemical 
potential $X\equiv \mu r^2$ that behaves as a negative constant and, for small perturbations, is related to the mean-field
solution by
\begin{equation}\label{ChemPot}
	X= -\frac{1}{1-\frac12 \varphi}\simeq -1 - \frac12 \varphi + \ldots \quad.\end{equation}
Notice that we are retrieving the solution found in the perturbative analysis when $\varphi$ plays the role of the 
integration constant $-A$.

Before moving on, it is mandatory to make a comment on the covariant conservation of our equations. At the end of section 2.1
we have pointed out that the diffeomorphism invariance entails a differential equation for the chemical
potential, namely (\ref{DiffEqChemPot}). The covariant conservation of the Einstein tensor implies automatically 
the same for the LHS of our equations of motion
\begin{equation}
	\mu_{,\beta}\bigl(h_\alpha{}^\beta-h_{\alpha\sigma}h^{\sigma\beta}+\tfrac14h^2\delta_\alpha^\beta\bigr)+\mu\bigl(h_\alpha{}^\beta-h_{\alpha\sigma}h^{\sigma\beta}+\tfrac14h^2\delta_\alpha^\beta\bigr){}_{;\beta}=0\,.\end{equation}
The latter equation is a set of 4 equations, $\alpha=t,r,\theta,\phi$, but only one is non trivial; this equation 
is the radial one, $\alpha=r$. When the perturbations are equal and constant, i.e. $h_t{}^t=h_r{}^r=\varphi$, the general 
covariance condition yields the following differential equation for the chemical potential $\mu$
\begin{equation}
	\biggl(\varphi-\frac{\varphi^2}{2}\biggr)\biggl(\partial_r\mu+\frac{2\mu}{r}\biggr)=0\,.\end{equation}
The integration is direct and the only degree of freedom is a boundary condition when integrating the differential 
equation that governs the chemical potential
\begin{equation}
	\mu=\mu_0\,r^{-2}\qquad\Longrightarrow\qquad
	X=\mu_0\,.\end{equation}
This value should coincide with the value of our negative and constant dimensionless chemical potential in (\ref{ChemPot}); 
this is $X=\mu_0=-1-\tfrac12\varphi$.

The qualitative result is the existence of a normalizable solution that can be interpreted as the collective wave 
function of a graviton condensate. A unique relation is obtained between this (constant) wave function and a (also constant) 
dimensionless chemical potential.

\section{Connection with previous proposals.}
It is immediate to see that the solution found for the wave function is of finite norm. Taking into account that 
the perturbation $h_{\alpha\beta}$ is null from the event's horizon onwards, the endpoint on the integration limit 
can be fixed at the Schwarzschild radius. This way, the integral is
\begin{equation}\label{Integralh2}
	\int_0^{\infty} d^3x \sqrt{-g} {h_\alpha}^\beta{h_\beta}^\alpha =4\pi\int_0^{r_s} dr\, r^2 \frac{h_t{}^{t\,2} 
+ h_r{}^{r\,2}}{\sqrt{(1-h_t{}^t)(1-h_r{}^r)}}= 4\pi\,{r_s}^3\,\frac{2\varphi^2 \, }{3\,(1-\varphi)}\end{equation}
and states that the integral of the square modulus of the wave function has a constant value. The volume element 
for the full metric $d^3x\sqrt{-g}=drd\Omega\,r^2\sin\theta/(1-\varphi)$ has been used. 
The fact that this magnitude is constant automatically ensures a constant behaviour for the probability density 
of the wave function defined in (\ref{Density}), as ${h_\alpha}^\beta{h_\beta}^\alpha=h_{\alpha\beta}h^{\alpha\beta}$.
Retrieving the missing constants, we can relate the latter quantity in (\ref{Integralh2}) to the integral of the 
density $\rho_{\hat h}$. Then, we are able to compute the total number of gravitons of the condensate. From the
arguments in section 2
\begin{equation}\label{mistery}
	N=\frac{8\pi}{3}\,{M_P}^2\, \frac{\varphi^2}{\,(1-\varphi)}\,{r_s}^2 \qquad\Longrightarrow\qquad r_s= \sqrt{\frac{3\,(1-\varphi)}{8\pi\varphi^2}}  \sqrt{N} L_P \,.\end{equation}
Here again the upper limit for the wave function enters explicitly; if $\varphi\rightarrow1$, then $N\rightarrow\infty$ and the metric becomes singular. 
At this point we should attempt to make contact with the results of \cite{DG1,DG2}. Under the maximum packaging 
condition $\lambda= r_s$ our previous relation agrees nicely with their proposal. The rest of relations of their work 
can be basically derived from this. 

Possibly our more striking results are that the dimensionless chemical potential $X= \mu(r) r^2$ stays constant and 
non-zero throughout the interior of the BH, and that so does the quantity $h_\alpha\,^\alpha=\varphi$ previously 
defined and entirely determined by the value of the dimensionless chemical 
potential $X$. Therefore, it is totally natural to interpret $X$ as the variable
conjugate to $N$, the number of gravitons.

As seen above the dimensionless chemical potential has a rather peculiar behaviour. As $X=\mu\,r^2$ is a constant 
function, then $\mu\propto 1/r^2$ and it is {\em not null} over the event horizon. Outside it appears to be exactly zero.
Let us now for a moment forget about the geometrical interpretation of BH physics and let us treat the problem as a 
collective many body phenomenon.  It is clear why gravitons are trapped behind the horizon: the jump 
of the chemical potential at $r=r_s$ would prevent 
the `particles' inside to reaching infinity. From this point of view it is quite natural to have a lower chemical potential
inside the horizon than outside (where is obviously zero) as otherwise the configuration would be thermodynamically unstable.
In the present solution particles (`gravitons' in our case) cannot escape.

However this is not completely true as the picture itself suggests that 
{\em one} of the modes can scape at a time without paying any
energy penalty if the maximum packaging condition is verified. 
Let us do a semiclassical calculation inspired by this picture; using $M\sim M_P\sqrt N$:
\begin{equation}
	\frac{dM}{dt}\simeq  M_P \frac{1}{2\sqrt N} \frac{dN}{dt}= \frac{1}{2r_s}\frac{dN}{dt}\,.\end{equation}
To estimate $dN/dt$ (which is negative) we can use geometrical arguments to determine the flux. If we assume that for
a given value of $r_s$ only one mode can get out (as hypothesized above) and that propagation takes place at the 
speed of light, elementary considerations\footnote{To determine the rate of variation of $N$ we have to multiply the surface 
($4\pi r_s^2$) times the flux; i.e. the density of the mode times the velocity, assumed to be $c=1$ in our units. Since
the density of the mode is constant in the interior, it is just $3/4\pi r_s^3$.} lead to
\begin{equation}
	\frac{dM}{dt}\simeq-\frac{3}{2}\frac{1}{r_s^2}\,.\end{equation}
This agrees with the results of \cite{DG2} ---for instance Eq. (35)--- and yields $T\simeq 1/r_s$.
Within this picture, several questions concerning the long-standing issue of loss of information 
may arise as the outcoming 
state looks thermal \cite{ThermalRadiation} but apparently is not; or at least not totally so. However we shall 
refrain of dwelling on this any further at this point.

The main result from the previous rather detailed analysis is that the BH is able to sustain a graviton
BEC and surely similar BECs made of other quanta\cite{das}. But is that condensate really present? Our results 
do not answer that question, but if we reflect on the case where the limit $N\to 0$ is taken, without 
disturbing the BH geometry (i.e. keeping $r_s$ constant) this requires taking $\varphi\to 0$ in a way that
the ratio $\sqrt{N}/|\varphi|$ is fixed. Then one gets $X=-1$. This value appears to be universal and
independent of any hypothesis. The metric is 100\% Schwarzschild everywhere. We conclude that a BH produces necessarily
a (trapping) non-zero chemical potential when the physical system is expressed in terms of the 
grand-canonical ensemble. Outside the BH, $X=0$ ($\mu=0$).

Another way of reaching this conclusion is by taking a closer look to our exact equations in the previous section.
If one sets $X=0$ then necessarily $\varphi=0$ and one gets the classical
Schwarzschild solution everywhere, but the reverse is not true. One can have $\varphi=0$ but
this does not imply $X=0$. Let us emphasize that these results go beyond the second order
perturbative expansion used in parts of this article.

As gravitons cross the horizon and are trapped by the BH classical gravitational 
field they eventually thermalize and form a BEC. The eventual energy surplus generated in this process is used to increase the 
mass and therefore the Schwarzschild radius of the BH. $\varphi$ is now non-zero; it is directly proportional 
to $\sqrt{N}$, the number of gravitons,
and the dimensionless chemical potential departs from the value $X=-1$, presumably increasing it in modulus. As soon as $\varphi\neq 0$
the metric inside the BH is not anymore Schwarzschild (but continues to be Schwarzschild outside). Note that the metric
is destabilized and becomes singular for $\varphi=1$, so surely this is an upper limit where $N\rightarrow\infty$.  

Yet another possible interpretation, that we disfavour, could be the following. Each BH has associated 
a given constant value of $X$, hence of $\varphi$. Then Eqs. (\ref{mistery}) would imply that after 
the emission of
each graviton, the value of $r_s$ is readjusted. The problem with this interpretation is that it would
require a new dimensionless magnitude ($X$) to characterize a Schwarzschild BH; something that most BH 
practitioners would probably find hard to accept.

\section{Conclusions and outlook}
We shall now conclude. The purpose of this work was to have some insights in the reformulation of a quantum theory 
of black holes in the language of condensed matter physics. The key point of the theory is to identify the black hole 
with a Bose-Einstein condensate of gravitons.

We have conjectured the set of equations that play the role of the Gross-Pitaevskii non-linear equation; they are derived from
the Einstein-Hilbert Lagrangian after adding a chemical potential-like term. We have used a number of different techniques 
to analyze these equations when the perturbation (i.e. the tentative condensate) has spherical symmetry. The
equations appear at first sight rather intractable, but by doing a perturbative analysis around the 
BH Schwarzschild metric at quadratic order (i.e. including the leading non-linearities) we found that the chemical 
potential necessarily vanishes in the exterior of the BH. On the contrary, in the interior we have found two sets of solutions, 
one of them has to be discarded as producing a non-normalizable result. The other
one leads to a non-zero chemical potential in the interior of the BH that behaves as $1/r^2$. Therefore, there is a finite jump
on the chemical potential at the BH horizon. Surprisingly ---or maybe not so--- this solution modifies the coefficient of the $tt$ and $rr$ terms in the Schwarzschild 
solution, but not its functional form. Of course if there is no chemical potential at all, the modification vanishes,
in accordance with well known theorems. However, if the former is non zero, the modification affecting the metric is also necessarily non-zero.

The perturbative analysis triggers a unique physical solution for the non-perturbative (exact) theory. This solution is characterized by a constant density of the wave function for the condensate. 
From the existence and knowledge of this solution, an unambiguous relation between the number of gravitons and the geometric properties of the BH is obtained. Hence, we find an expression for the Schwarzschild radius that involves an a priori independent and tuneable parameter, the dimensionless chemical potential $X$ (related to the mean-field wave function of the condensate $\varphi$). We find this somewhat strange as this would be a new black hole parameter. Therefore we favour the universal value $X=-1$ as
discussed in the text. From this expression for the Schwarzschild radius, most relations obtained in \cite{DG1,DG2} can be rederived.

As should be obvious to the reader who has followed our discussion, our approach is somewhat different from the one developed
in the initial papers by Dvali, G\'omez and coworkers. We assume from the start the existence of a classical geometry background
that acts as confining potential for the condensate. The fact that the functional form of the metric perturbation induced by
the condensate is exactly the same as the original background, of course gives a lot of credence to the possibility of deriving
the latter from the former in a sort of self-consistent derivation. We have not explored this possibility in detail yet.  

It is quite plausible that one could entertain the presence of condensates of other quantum fields inside the BH horizon (why
only gravitons?). While we do not expect much of a conceptual difference, it would be 
very interesting to see the similarities and differences with the case of quantum gravitons.

\section*{Acknowledgments}
We acknowledge the financial support of projects
FPA2013-46570-C2-1-P (MINECO) and 2014SGR104 (Generalitat de Catalunya). This research was supported in part though the E.U. EPLANET
exchange program FP7-PEOPLE-2009-IRSES, project number 246806. The work of J.A. is partially supported by grants
Fondecyt 1150390 and CONICYT-PIA-ACT14177 (Government of Chile).

\end{document}